\documentclass[letter]{article}

\usepackage{graphicx}
\usepackage{amssymb}
\usepackage{bm}                
\usepackage{mathrsfs}      
\usepackage{amsmath}
\usepackage{amsthm}

\usepackage{natbib}

\newcommand{\nn}{\nonumber}
\newcommand{\Ind}{1\!\mathrm{l}}
\newtheorem{corollary}{Corollary}

\begin{document}
\title{Hellinger Distance and Bayesian Non-Parametrics: \\ Hierarchical Models for Robust and Efficient Bayesian Inference}

\author{{Yuefeng} {Wu}  and {Giles} {Hooker}\\
University of Missouri St. Louis and Cornell University}

\date{}
\maketitle

\newtheorem{theorem}{Theorem}
\newtheorem{lemma}{Lemma}
\newtheorem{remark}{Remark}
\newtheorem{definition}{Definition}
\newtheorem{proposition}{Proposition}
\newtheorem{ex}{}

\setlength{\baselineskip}{1.5\baselineskip}

\begin{abstract}

This paper introduces a hierarchical framework to incorporate Hellinger distance methods into Bayesian analysis. We propose to modify a prior over non-parametric densities with the exponential of twice the Hellinger distance between a candidate and a parametric density. By incorporating a prior over the parameters of the second density, we arrive at a hierarchical model in which a non-parametric model is placed between parameters and the data.  The parameters of the family can then be estimated as hyperparameters in the model. In frequentist estimation, minimizing the Hellinger distance between a kernel density estimate and a parametric family has been shown to produce estimators that are both robust to outliers and statistically efficient when the parametric model is correct. In this paper, we demonstrate that the same results are applicable when a non-parametric Bayes density estimate replaces the kernel density estimate. We then demonstrate that robustness and efficiency also hold for the proposed hierarchical model. The finite-sample behavior of the resulting estimates is investigated by simulation and on real world data.
\end{abstract}





\section{Introduction}
This paper develops Bayesian analogs of Hellinger distance methods through the use of a hierarchical formulation. In particular, we aim to produce methods that enable a Bayesian analysis to be both robust to unusual values in the data and to retain their precision when a proposed parametric model is correct.
All statistical models include assumptions which may or may not be true for given data set. Robustness is a desired property in which a {statistical} procedure is relatively insensitive to the deviations from these assumptions. For frequentist inference, concerns are largely associated with distributional robustness: the shape of  the true underlying distribution deviates slightly from the assumed model. Usually, this deviation represents the situation where there are some outliers in the observed data set{; see \citet{Huber2004} for example}.   For Bayesian procedures, the deviations may come from the model, prior distribution, or utility function or some combination thereof. Much of the literature on Bayesian robustness has been concerned with the prior distribution or utility function.  By contrast, the focus of this paper is  robustness with respect to outliers in a Bayesian context.  However, there has been little study of this form of robustness for  Bayesian models. For example, we know  {Bayesian models with heavy tailed data distributions} are robust with respect to outliers for the case of one single location parameter estimated by many observations; however, we have only a few sparse results for the case of models with more than one parameter {and few results for hierarchical mdoels}.  The hierarchical method we propose, and the study of its robustness properties, {will provide an alternative means of making any data distribution robust to outliers}.


Throughout this paper, we suppose that we have a parametric family of univariate data generation models $\mathscr F=\{f_{\theta}:\theta\in\Theta\}$ for some parameter space $\Theta$. We are given the task of estimating $\theta_0 \in \Theta$ from univariate i.i.d. data $X_1,\ldots,X_n$ where we assume each $X_i$ has density $f_{\theta_0}$ for some true parameter value $\theta_0$. The statistical properties of our proposed methods for accomplishing this will be examined below. Throughout, convergence results are given with respect to the measure $P_{\theta_0}^{\infty}$ -- the distribution of i.i.d. sequences generated according to $f_{\theta_0}$.  The generalization to a generating density $g \notin \mathscr F$ can be made in a straightforward manner, but at the cost of further mathematical complexity and we do not pursue this here.


Within the frequentist literature,  minimum Hellinger distance estimates proceed by first estimating a kernel density $\hat g_n(x)$ and then choosing $\theta$ to minimize the Hellinger distance $\int(\sqrt{f_{\theta}(x)} - \sqrt{\hat g_n(x)})^2 dx$.  The minimum Hellinger distance estimator was shown in \cite{Beran77} to have the remarkable properties of both being robust to outliers and statistically efficient -- in the sense of asymptotically attaining the information bound -- when the data are generated from $f_{\theta_0}$. These methods have been generalized to a class of minimum disparity estimations, which have been studied since then, (eg. \cite{BasuLindsay94,BSV97,PakBasu98,ParkBasu04} and \cite{Lindsay94}). In this paper, {only consider} Hellinger distance in order to simplify the mathematical exposition; the extension to more general disparity methods can be made following a similar developments to those in \cite{ParkBasu04} and \cite{BSV97}.

Recent methodology proposed in \cite{Hooker11}, suggested the use of disparity-based methods  within Bayesian inference via the construction of a ``disparity likelihood'' by replacing the likelihood function when calculating the Bayesian posterior distribution; {they} demonstrated that the resulting expected {\em a posteriori} estimators retain the frequentist properties studied above. However, these methods first obtain kernel nonparametric density estimates from data, and then calculate the disparity between the estimated density function and the corresponding density functions in the parametric family. In this paper, we propose  the use of Bayesian non-parametric methods to marginalize a posterior distribution for the parameters given a non-parametric density.
This represents a natural incorporation of disparities into Bayesian analysis: a non-parametric representation of the distribution is placed between the parametric model and the data and we tie these together using a disparity.  We show that this approach is more robust than usual Bayesian methods and  demonstrate that  the expected {\em a posteriori} estimators of $\theta$ retain asymptotic efficiency, hence the precision of the estimate is maintained.

In this paper, we will study the use of Dirichlet normal mixture prior for non-parametric densities within our methods. These priors were introduced by \cite{Lo84} (see also \cite{Ghorai82}), who obtained expressions for the resulting posterior and predictive distributions. We use normal density $\phi(x,\omega)$ as the kernel function used in the mixture model. Let $g_P=\int_{\Omega} \phi(x,\omega)dP(\omega)$ for any probability $P$ on $\Omega$, then the Dirichlet process prior $\Pi$ on the space of probability measures on $\Omega$ gives rise to a prior on densities via the map $P\mapsto g_P$. The asymptotic properties of such models have been studied by \cite{ghosal99, ghosal00}, \cite{ghosal01,ghosal07} and \cite{wu08,wu10}. 

To examine the asymptotic properties of our proposed methods, we begin by  examining {one-step} minimum Hellinger distance methods in which the kernel density estimate is replaced with a nonparametric Bayes estimator. {The sampling properties of these estimates are inherited implicitly from those of the non-parametric density estimator.} The asymptotic properties of the application of minimum Hellinger distance methods with {Bayesian}  density estimators remains an open question.  We define three possible means of combining a Bayesian posterior non-parametric density estimate with Hellinger distance, which we call one-step methods. These results will {then} be used for establish the efficiency  of the proposed hierarchical {formulation}.

By using the asymptotic results for Dirichlet normal mixture priors, the properties of the one step methods,  such as consistency and efficiency can be obtained in straightforward manner. We briefly discuss these procedures in Section 2. A hierarchical model is introduced in Section 3 where we establish consistency and efficiency for this estimator, too.  Section 4 studies the robustness of the procedures and Section 5 reports the simulation performance of these methods with modest sample sizes.

\section{One-step methods}

In this section, we examine the asymptotic properties of replacing kernel density estimates with Bayesian non-parametric density estimates within minimum Hellinger distance estimation. For simplicity, we use $\mathbb X_n$ to denote the observations $X_1,\ldots,X_n$.  We assume that $f(\cdot,\theta)$ is continuous in $\theta$  and possesses at least 3rd derivatives. {We also make the identifiability assumption}
\begin{itemize}
\item[I1] \label{ident} $\theta_0 \in \Theta$ is identifiable in the sense that for every $\delta$ there exists $\delta^*$ such that
\[
|\theta - \theta_0| > \delta \Rightarrow D_H(f_{\theta},f_{\theta_0}) > \delta^*.
\]
\end{itemize}

To estimate $\theta$ given $\mathbb X_n$, we first introduce a Bayesian non-parametric estimate of the density.
 Let $\mathscr G$ be the space of all probability density functions on $\mathbb R$ with respect to Lebesgue measure {and define a topology on $\mathscr G$ given by Hellinger distance}. Let $\Pi$ denote a prior on $\mathscr G$.  For any measurable subset $B\subset \mathscr G$, the posterior probability of $g\in B$ given $\mathbb X_n$ is
$$
\Pi(B|\mathbb X_n)=\frac{\int_B\prod_{i=1}^ng(X_i)d\Pi(g)}{\int_{\mathscr G}\prod_{i=1}^ng(X_i)d\Pi(g)}.
$$

The squared Hellinger distance between $g$ and $f_{\theta}$ is
\begin{equation}
{\rm D_H}(g,\theta):=\int \left(f^{1/2}(\cdot, \theta)-g^{1/2}(\cdot)\right )^2d\lambda,
 \end{equation}
 where $\lambda$ is the Lebesgue measure on $\mathbb R$. A functional $T$ on $\mathscr G$ is then defined as the following:
 for every $g\in \mathscr G$,
$$
\|f^{1/2}_{T(g)}-g^{1/2}\|^2=\min_{t\in\Theta}\|f_t^{1/2}-g^{1/2}\|^2,
$$
where $\|\cdot\|$ denotes the $L_2$ metric. For the existence and continuity of $T(g)$,  refer to Theorem 1 in \cite{Beran77}.

We propose the following three estimators for $\theta$:

1. Minimum Hellinger distance estimator:
\begin{equation}\label{eq:model3}
\hat \theta_1={\rm argmin}\ D_H \left (\theta,\int g \Pi(dg|\mathbb X_n) \right ).
\end{equation}
This estimator  just replaces the kernel density estimate in the classical minimum Hellinger distance method by the posterior expectation of the density function. Let $g_n^*$ denote $\int g \Pi(dg|\mathbb X_n)$, we can write $\hat \theta_1$ as $T(g_n^*)$.

2. Minimum (posterior) expected Hellinger distance estimator:
\begin{equation}\label{eq:model1}
\hat\theta_2={\rm argmin}_{\theta}\int {\rm D_H}(g,\theta)\Pi(dg|\mathbb X_n).
\end{equation}
This minimizes the expectation of the Hellinger distance between the density function from the parametric family and any density function on $\mathbb R$ with respect to the posterior distribution of the density functions.

3. Minimum (posterior) probability estimator:
\begin{equation}\label{eq:model2}
\hat{\theta}_3 (\epsilon_n) = {\rm argmin}\  \Pi \{ g: {\rm D_H}(g,\theta)>\epsilon_n|\mathbb X_n \},
\end{equation}
for some $\epsilon_n$.
This estimator finds the $\hat {\theta}_3$  that minimizes the posterior probability of the density functions that are at least $\epsilon_n$ away from it in $D_H$. This estimator is constructed to reflect the way in which we study the convergence rate of the posterior distribution. The rate of convergence is {defined by choosing} $\epsilon_n \to 0$ such that $\Pi\{ g: D_H(g,f_{\theta_0})>\epsilon_n|\mathbb X_n\}\to 0$. In \cite{ghosal00}, a similar estimator was studied in a purely nonparametric Bayesian model.

We examine the large sample behavior of these estimators in the following theorem and show  that all these estimators give estimates that converge to the true value in probability.

\begin{theorem}\label{thm:1}

If for any given $\epsilon>0$, $\Pi\{g: D_H(g,f_{\theta_0})>\epsilon|\mathbb X_n  \}\to 0$ in probability, under assumption I1,
\begin{itemize}
\item [1.] $\| g_n ^{*1/2}-f_{\theta_0}^{1/2}\|^2\to 0$ in probability, and if $T$  is continuous at $g_0$ in Hellinger distance, then $T(g_n^{*})\to T(f_{\theta_0})$ in probability, and hence $\hat\theta_1 \to \theta_0$ in probability;

\item [2.] for any $\theta_0\in \Theta$, we have that $\hat\theta_2 \to \theta_0$ in probability;
\item [3.] for any $\theta_0\in \Theta$, and if $T$  is continuous at $g_0$ in Hellinger distance, and further if there exist $\epsilon_n\downarrow 0$, $\Pi\{g: D_H(g,f_{\theta_0})>\epsilon_n|\mathbb X_n  \}\to 0$ in probability,  we have that $\hat\theta_3(\epsilon_n) \to \theta_0$ in probability.
\end{itemize}
 \end{theorem}

{\sc Proof.}
Because that the squared Hellinger distance is bounded from above by 2, it is easy to see that $\| g_n ^{*1/2}-g_0^{1/2}\|\to 0$ in probability since for any given $\epsilon>0$, $\Pi\{g: D_H(g,f_{\theta_0})>\epsilon|\mathbb X_n  \}\to 0$ in probability. Then by Theorem 2.2 in \cite{Cheng06}, part 1 of this theorem follows.

 For estimator $\hat \theta_2$, we have that $\int D_H(f_{\theta_0}, g)d\Pi(g|\mathbb X_n)<\epsilon$ for any given $\epsilon>0$, since for any given $\epsilon>0$, $\Pi\{g: D_H(g,f_{\theta_0})>\epsilon|\mathbb X_n  \}\to 0$ in probability. By the definition of $\hat \theta_2$, we have that $\int D_H(f_{\hat \theta_2}, g)d\Pi(g|\mathbb X_n) \leq \epsilon $, which implies that $D_H(\hat \theta_2,\theta_0)\leq (2\sqrt\epsilon)^2=4\epsilon$ and hence part 2 of the theorem holds.

 For estimator $\hat \theta_3(\epsilon_n)$, by definition, we have that $\Pi\{g:D_H(g,f_{\hat\theta_1})\geq \epsilon_n|\mathbb X_n\} \geq \Pi\{g:D_H(g,f_{\hat\theta_3}(\epsilon_n))\geq \epsilon_n|\mathbb X_n\}$. By the result in part 1, we have that for any given $\epsilon>0$, $P(D_H(f_{\hat \theta _3},f_{  \theta_0})\leq (3\sqrt\epsilon)^2=9\epsilon) \rightarrow 1$, and hence $\hat \theta_3(\epsilon_n) \to \theta_0$ in probability.

$\Box$

\begin{remark}
The $\epsilon_n$ used above is called the rate of the convergence in the context of Bayesian nonparametric density estimation. Usually, if the true density $f_{\theta_0}$ is of the form of the mixture of the kernel functions, then the best rate is $\epsilon_n=\log n/\sqrt n$, see \cite{walker07}  for more details. If  the true density $f_{\theta_0}$ is smooth and has second derivatives, then as showed in \cite{ghosal07}, the best rate is $n^{-2/5(\log n)^{4/5}}$.
\end{remark}

\begin{remark}
These estimates above assume that $f_{\theta_0}$ is the data generating distribution. When the data are generated from a density $g_0$ not in the parametric family $\mathcal{F}$, {similar}  arguments {yield}  {the consistency of} $\theta_0$ {defined}  to minimize $D_H(g_0,f_{\theta})$.
\end{remark}

If we replace the $\epsilon$ in Theorem \ref{thm:1} by $\epsilon_n$, the convergence rate of nonparametric Bayesian density estimation will only give  lower bounds of the convergence rates of these three estimators. In the following theorems we show that if the Bayesian density estimate $g_n^*$ satisfies that for any $\sigma\in L_2$ with $\sigma \perp f_{\theta_0}^{\frac{1}{2}}$ under the usual inner product, the limit distribution of $n^{\frac{1}{2}}\int \sigma(x)[g^{*\frac{1}{2}}_n(x)-f_{\theta_0}^{\frac{1}{2}}(x)]dx$ is $\rm {Norm}(0,\|\sigma\|_{f_{\theta_0}}^2/4)$, where $\|\sigma\|_g^2$ denotes $\int \sigma^2(x)g(x)dx$, then $\hat{\theta}_1$ is asymptotically normally distributed with the variance equivalent to the inverse of the Fisher's information $I(\theta_0)^{-1}$.


We  assume that for specified $t\in \Theta\subset \mathbb R^p$, there exist a $p \times 1$ vector $\dot f^{1/2}_t = \frac{df^{1/2}_t(x)}{dt}$ with components in $L_2$ and a $p\times p$ matrix $\frac{d^2 f^{1/2}_t(x)}{dt^2}$ with components in $L_2$ such that for every $p \times 1$ real vector $l$ of unit Euclidean length and for every scalar $a$ in a neighborhood of zero,
\begin{equation}\label{eq:25}f^{1/2}_{t+al}(x)=f^{1/2}_t(x)+al^T\frac{df^{1/2}_t(x)}{dt}+al^Tu_a(x),\end{equation}
and
\begin{equation}\label{eq:26}\dot f^{1/2}_{t+al}(x)=\frac{df^{1/2}_t(x)}{dt}+al^T\frac{d^2f^{1/2}_t(x)}{dt^2}+al^Tv_a(x),\end{equation}
where $u_a(x)$ is $p\times 1$, $v_a(x)$ is $p\times p$, and the components of $u_a$ and of $v_a$ individually tend to zero in $L_2$ as $a\to 0$. {This}  assumption makes $T$ {a}  differentiable functional, which  is fundamental for the rest of this paper. Some convenient sufficient conditions for (\ref{eq:25}) and (\ref{eq:26}) were given by Lemma 1 and Lemma 2 in \cite{Beran77}.

\begin{theorem}\label{thm:2}
Suppose that
 \begin{itemize}
 \item[A1]
 Expression (\ref{eq:25}) and (\ref{eq:26}) hold for every $t\in \Theta$, $T(g)$ exists, is unique, and lies in $\mbox{\rm int}(\Theta)$, $\int \frac{d^2f^{1/2}_{T(g)}(x)}{dt^2}g^{1/2}(x)dx$ is a nonsingular matrix, and the functional $T$ is continuous at $g$ in the Hellinger topology,
\item[A2]
For any $\sigma$, such that $\sigma\in L_2$ and $\sigma \perp g^{\frac{1}{2}}$, the limit distribution of $n^{\frac{1}{2}}\int \sigma(x)[g^{*\frac{1}{2}}_n(x)-g^{\frac{1}{2}}(x)]dx$ is $N(0,\|\sigma\|_g^2/4)$.
\end{itemize}
Then the limiting distribution of $n^{1/2}[T(g_n^{*\frac{1}{2}})-T(g)]$ under $g$ as $n\to \infty$ is $N \left (0,4^{-1}\int \rho _g(x)\rho^T_g(x)dx\right )$ where $$\rho_g(x)=\int
\left[\int\frac{d^2 f^{1/2}_{T( g)}(x)}{dt^2} g^{\frac{1}{2}}(x)dx \right]^{-1}\frac{d f^{1/2}_{T( g)}(x)}{dt}.$$ If $g=f_{\theta_0},$ the limiting distribution of $n^{1/2}[T(g^{*\frac{1}{2}})-T(g)]$ under $g$ as $n\to \infty$ is $$N \left (0,4^{-1}\left [\int \frac {df^{1/2}_{\theta_0}(x)}{dt}{\frac{d f^{1/2}_{\theta_0}(x)}{dt}}^Tdx\right ]^{-1}\right ) = N\left(0,I(\theta)^{-1}\right).$$
where $I(\theta)$ is the Fisher information for $\theta$ in the family $f_\theta$.
\end{theorem}

{\sc Proof.}
 When Condition A1 holds, based on Theorem 2 in \cite{Beran77} and its proof, we have the following:
\begin{eqnarray}
T(g^*_n) \int
&=&T( g)+\rho_g(x)
[ g_n^{*\frac{1}{2}}(x)- g^{\frac{1}{2}}(x)]dx\nn\\
&&+V_n\int \frac{d f^{1/2}_{T( g)}(x)}{dt}[ g_n^{*\frac{1}{2}}(x)- g^{\frac{1}{2}}(x)]dx,
\end{eqnarray}
where $V_n\to 0$ in probability, $\rho_g \perp g$  and $\frac{d f^{1/2}_{T( g)}(x)}{dt}\perp g(x)$. Then by Condition A2 and $\left[\rho_g(x)+V_n\frac{d f^{1/2}_{T( g)}(x)}{dt}\right]\left[\rho_g(x)+V_n\frac{d f^{1/2}_{T( g)}(x)}{dt}\right]^T \to \int \rho_g(x)\rho_g^T(x)dx$ in probability as $n\to\infty$,  the proof is completed.
$\Box$\\

The proof of Theorem \ref{thm:2} heavily relies on assumption A2, which may or may not hold for general Bayesian nonparametric density estimates. The following lemma gives sufficient conditions on Bayesian nonparametric density estimates, under which condition A2 holds.



 Let $\hat g_n$ be a kernel density estimator
 $$
 \hat g_n(x)=(nc_ns_n)^{-1}\sum_{i=1}^nk[(c_ns_n)^{-1}(x-X_i)],
 $$
where $c_n$ is a sequence of constants converging to zero at an appropriate rate, $s_n=s_n(X_1, \ldots,X_n)$ is a robust scale estimator, and $k$ is a smooth density on the real line.

 Let $\tilde g_n$ be
 $$
 \tilde g_n(x)=(nc_ns_n)^{-1}\int k[(c_ns_n)^{-1}(x-y)]dG(y).
 $$

Let $G_n$ denote the empirical cdf of $(X_1,X_2,\ldots,X_n)$ and
$\tilde G^*_n$ be the cdf corresponding to $\tilde g^*_n$. We define $ \tilde g^*_n$ be
 $$
 \tilde g^*_n(x)=(nc_ns_n)^{-1}\int k[(c_ns_n)^{-1}(x-y)]d\tilde G^*_n(y),
 $$
and
\begin{equation}\label{eq:bn}
B_n(x)=n^{\frac{1}{2}}[\tilde G^*_n-G].
\end{equation}




\begin{lemma} \label{lem:1}
 Suppose that
 $$
 \int \frac{\sigma(x)}{2g^{\frac{1}{2}}(x)}dB_n(x)\sim \textrm{Norm}(0,\|\sigma\|^2/4),
 $$
 where the $B_n(x)$ is defined in (\ref{eq:bn}) corresponding to the Bayesian nonparametric density estimate $g^*_n(x)$, the concentration rate for  $g^*_n(x)$ is $o(n^{-\frac{1}{4}})$, and  condition A1 in Theorem \ref{thm:2} holds, and
\begin{itemize}
\item[b1.] $k$ is symmetric about $0$ and compact support on $K$,
\item[b2.] $k$ is twice absolutely continuous; $k''$ is bounded,
\item[b3.] $g>0$ on $K$, $g$ is twice absolutely continuous and $g''$ is bounded,
\item[b4.] $\lim_{n\to \infty}n^{1/2}c_n=\infty$, $\lim_{n\to \infty}n^{1/2}c_n^2=0$, and
\item[b5.] there exists a positive finite constant $s$ depending on $g$ such that $n^{1/2}(s_n-s)$ is bounded in probability.
\end{itemize}
Then Condition A2 of Theorem \ref{thm:2} is satisfied.
\end{lemma}

{\sc Proof}. 

For $b\geq 0, a>0$, we have
\begin{equation}
b^{\frac{1}{2}}-a^{\frac{1}{2}}=(b-a)/(2a^{\frac{1}{2}})
-(b-a)^2/[2a^{\frac{1}{2}}(b^{\frac{1}{2}}+a^{\frac{1}{2}})^2].
\end{equation}
Thus,
\begin{equation}
n^{\frac{1}{2}}\int \sigma(x)[g^{*\frac{1}{2}}_n-g^{\frac{1}{2}}(x)]dx=n^{\frac{1}{2}}\int_K \sigma(x)[g^{*}_n-g(x)]/(2g^{\frac{1}{2}}(x))dx+R_n
\end{equation}
where, for $\delta=min_{x\in K} g(x)>0$.
\begin{eqnarray*}
|R_n|&\leq& n^{\frac{1}{2}}\int_K |\sigma(x)|[g^{*}_n-g(x)]^2/(2g^{\frac{3}{2}}(x))dx\\
     &\leq& 2\delta^{-\frac{3}{2}}n^{\frac{1}{2}}\int |\sigma(x)|[g^{*}_n- g(x)]^2dx\to 0,
\end{eqnarray*}
in probability, due to the condition on the concentration rate of $g^*_n$.

Let $\psi(x)=\sigma(x)/(2g^{\frac{1}{2}}(x))$  and write
\begin{eqnarray*}
\lefteqn{n^{\frac{1}{2}}\int \psi(x)[g^*_n(x)-g(x)]dx}\\
&=&n^{\frac{1}{2}}\int \psi(x)[g^*_n(x)-\tilde g^*_n(x)]dx+n^{\frac{1}{2}}\int \psi(x)[\tilde g^*_n(x)-g(x)]dx\\
&=&V_{1n}+V_{2n}, \quad \mbox{say.}
\end{eqnarray*}
Replacing $G$ by $G^*_n$, by the same argument in \cite{Beran77} for $U_{3n}$ defined in \cite{Beran77}, we have that $V_{1n}$ converges to 0 as $n\to \infty$.
Let
\begin{eqnarray*}
T_{1n}&=&n^{-\frac{1}{2}}(c_ns)^{-1}\int k[(c_ns)^{-1}(x-y)]dB_n(y),\\
T_{2n}&=&n^{-\frac{1}{2}}\int_{c_ns}^{c_ns_n} t^{-2}dt \int B_n(x-tz)[2k'(z)+zk''(z)]dz.
\end{eqnarray*}
We can write
\begin{eqnarray*}
\lefteqn{V_{2n}}\\
&=&n^{\frac{1}{2}}\int \psi(x)T_{1n}(x)dx+n^{\frac{1}{2}}\int \psi(x)T_{2n}(x)dx+n^{\frac{1}{2}}\int \psi(x)[\tilde g_n(x)-g(x)]dx.
\end{eqnarray*}
It is easy to see that the second and third terms on the right hand side of the equation above both converge to $0$ by the same arguments in \cite{Beran77}. The first term there can be expressed as $\int B_n(y)\int \psi(y+c_nsz)k(z)dz$. Due to the condition on $B_n(x)$, the lemma follows. $\Box$\\

\begin{remark}
Lemma \ref{lem:1} is also useful for $g$ that is not compactly support. Refer to \cite{Hooker11} for more details.
\end{remark}

We define random histogram prior as follows: $h\sim \mu$; given $h$, choose $P$ on $\mathbb Z=\{0,\pm1,\pm2,\ldots\}$ with $P\sim \mathscr D_{\alpha h}$, where $\mathscr D_{\alpha h}$ is a Dirichlet process with parameter $\alpha h$; and $X_1,X_2,\ldots,X_n$ are, given $h, P$, i.i.d. $f_{h,p}$, where
$$
f_{h,p}=\sum_{-\infty}^{\infty} \frac{P(i)}{h}\Ind_{(ih,(i+1)h]}(x).
$$

\begin{remark}
Refer to \cite{ghosal00} for the approximation property of the uniform density function kernel, refer to \cite{wu08} for the concentration rate result, and refer to Chapter 5 in \cite{Ghosh2003} for the explicit expression of the random histogram density estimate, it is clear that the random histogram prior satisfies lemma \ref{lem:1}.
\end{remark}

\begin{remark}
Due to the flexibility of Bayesian nonparametric density estimation and the large size of the space of the density functions, it is difficult to obtain a general result for asymptotic normality. However, besides the random histogram prior studied above, it is not hard to see that if we can somehow control the tail property of the kernel functions, which are used for the often used Dirichlet mixture priors, we will have the same asymptotic normality. Also, Dirichlet mixture prior will give a ``parametric'' estimation if the base measure of Dirichlet process $\alpha(\mathbb R)\to 0$ and in this case, the asymptotic normality follows.
\end{remark}

In the following theorem, we give sufficient conditions under which $n^{1/2}(\hat {\theta}_2-\hat{\theta}_1)\to 0$ in probability, and hence the asymptotic normality of $\hat {\theta}_1$ implies the asymptotic normality of $\hat {\theta}_2$. In order to do so, we make the following assumption:
\begin{itemize}
\item[(B1)] $D_H(\theta,g)$ is thrice differentiable with respect to $\theta$ in a neighborhood $(\theta_0-\delta,\theta_0+\delta)$ for any $g$. If $\dot D_H, \ddot D_H$ and $\dddot D_H$ stand for the first, second and third derivatives then $E_{\theta_0}\dot D_H(\theta_0)$ and $E_{\theta_0}\ddot D_H(\theta_0)$ are both finite and
    $$
    \sup_{\theta_0-\delta,\theta_0+\delta}\left |\int \dddot D_H(\theta,g) \Pi(dg|x)\right |<M(x) \mbox {and }E_{\theta_0}M<\infty.
    $$
\end{itemize}
These conditions are more stringent than needed here, but they will be used in Theorem \ref{thm:h} below. Note that this condition only requires some regularity in the model $f_\theta$. We can now obtain

\begin{theorem} \label{thm:3}
Let $\Pi_n$ denote the posterior distribution of the Bayesian nonparametric density estimation and $g_n^*=\int gd\Pi_n$. If $n\int D_h(g,g_n^*)d\Pi_n\to 0$ in probability, then $n^{1/2}(\hat {\theta}_2-\hat{\theta}_1)\to 0$ in probability.
\end{theorem}
{\sc Proof}. By condition (B1), we have that the first order derivative exists and the second order derivatives are finite. Therefore, we only need to show that the difference between the target functions for the corresponding estimators is $o(n^{-1/2})$. Let $\Pi_n$ denote $\Pi(g|\mathbb X_n)$, we have that
\begin{eqnarray}\label{eq:n12}
\lefteqn{\left|D_H\left(\theta,\left(\int g d\Pi_n\right)\right)-\int D_H(\theta,g)d\Pi_n\right|}\\
&=&\left|2-2\int f_{\theta}^{1/2}\left(\int g d\Pi_n\right)^{1/2}dx-\int \left[2-2\int f_{\theta}^{1/2}g^{1/2}\right]d\Pi_n \right|\nn\\
&=&2\int \left\{\int f_{\theta}^{1/2} \left|g^{1/2}-\left(\int gd\Pi_n\right)^{1/2}\right|d\lambda    \right \}    d\Pi_n   \nn\\
&\leq& 2\int \sqrt{\int\left(g^{1/2}-\left(\int gd\Pi_n\right)^{1/2}\right)^2dx   }  \ \  d\Pi_n\nn\\
&=&2\int D_H(g,g_n^*)d\Pi_n\nn \\
& = & o_p(1/\sqrt{n}) \nn
\end{eqnarray}
by \eqref{eq:n12}.
$\Box$

As a consequence of Theorem \ref{thm:3} and the efficiency of $\hat{\theta}_1$ we have
\begin{corollary}
Under the conditions of Theorems \ref{thm:2} and \ref{thm:3}, $\sqrt n(\hat{\theta}_2-\theta_0)\stackrel{\mathscr D} \to N(0,1/I(\theta_0))$.
\end{corollary}

\section{Hierarchical Method}

In the previous section we examined three one-step minimum Hellinger distance estimates that incorporate a Bayesian non-parametric posterior distribution in place of a kernel density estimate. These methods have the advantage of providing an automatic method of bandwidth selection (when using appropriate priors), but the parameter estimates are post-hoc projections of the non-parametric posterior onto the parametric family $f_\theta$ rather than being included in a Bayesian analysis.

\cite{Hooker11} proposed the use of Hellinger distance and other disparities within Bayesian inference, but retained the kernel density estimate. This was accomplished by replacing the log likelilhood with $2n$ times the Hellinger distance between a parametric model and the kernel density estimate and was shown to provided both robustness and asymptotic efficiency in Bayesian estimates.   In this section, we provide a unified framework for disparity-based Bayesian methods in which the ``Hellinger likelihood'' of \cite{Hooker11} modifies the prior for the non-parametric density estimate. This creates a hierarchical model that can be viewed as placing a non-parametric estimate between the data and the proposed parametric model. While we treat the problem of estimation with i.i.d. data, this notion can be considerably expanded; see \cite{Hooker11} for more details.

Specifically, the model uses the Hellinger distance measure of the disparity between the nonparametric model and the parametric model  to construct the alternative of  the conditional likelihood  ${\rm Pr}(\theta|g)$, then integrating ${\rm Pr}(\theta|g){\rm Pr}(g|\mathbb X_n)$ with respect to $g$ gives the ``Hellinger posterior density'' of the parameter $\theta$, which completes the model.

Let $\pi(\theta)$ denote  the prior probability density function of $\theta$ and $\Pi(g)$ denote the prior probability distribution of the density function $g$.
We define the Hellinger posterior density function as
\begin{equation}\label{eq:model5}
\Pi^H(\theta,g|\mathbb X_n)= \frac{\pi(\theta)e^{-2nD_H(g,\theta)}}{\int\pi(\theta)e^{-2nD_H(g,\theta)}d\theta}\frac{\Pi(d g)g(\mathbb X_n)}{\int\Pi(d g)g(\mathbb X_n)}.
\end{equation}
The first term in the product is the Hellinger posterior of \cite{Hooker11} for fixed $g$. Here, we have used it to modify the prior over densities $g$. Under this model, the hierarchical Hellinger posterior density for $\theta$ is obtained by marginalizing over $g$:
\begin{equation}\label{eq:hp}
\Pi^H(\theta|\mathbb X_n)= \int_{\mathscr G}\Pi^H(\theta,g|\mathbb X_n)dg.
\end{equation}

First, we  show that under some sufficient conditions, the hierarchical model is consistent.
\begin{theorem} \label{thm:4}
Under the conditions of Theorem \ref{thm:1}, let $\pi$ be a prior on $\Theta$, $\Pi$ be a prior on $\mathscr G$ and $\pi(\cdot|g)$ and $\Pi(\cdot|\mathbb X_n)$ be the respective posteriors. Assume that $\theta_0$ is the true parameter. If
the $\theta_0$ belongs to the support of $\pi$, and the prior $\Pi$ yields a strongly consistent posterior $\Pi(\cdot|\mathbb X_n)$ at $f_{\theta_0}$, then the Hellinger Posterior as defined in (\ref{eq:hp}) degenerates to a point mass at $\theta_0$ almost surely as $n$ goes to infinity.
 \end{theorem}

{\sc Proof}.
Under the {identifiability} assumption I1
we have that
\begin{eqnarray}\label{eq:consistent}
\lefteqn{\Pi^H(|\theta-\theta_0|>\delta|\mathbb X_n)}\nn\\
&=&\int_{D_H(f_{\theta},f_{\theta_0})>\delta^*}\int_{\mathscr G} \frac{e^{-2nD_H(g,\theta)}\pi(\theta)}{\int e^{-2nD_H(g,\theta)} \pi(\theta)d\theta}d\Pi(g|\mathbb X_n) d\theta,
\end{eqnarray}
for some $\delta^*>0$. Let $\epsilon <\delta^*/4$, we have that (\ref{eq:consistent}) is less than
\begin{eqnarray}\label{eq:consistent1}
&& \int_{D_H(f_{\theta},f_{\theta_0})>\delta^*}\Big [ \int_{D_H(g,f_{\theta_0})\leq \epsilon/4}\frac{e^{-2nD_H(g,\theta)}\pi(\theta)}{\int e^{-2nD_H(g,\theta)}\pi(\theta)d\theta}d\Pi(g|\mathbb X_n)\nn\\
&&\qquad\qquad\quad+\pi(\theta)\Pi[D_H(g_n,f_{\theta_0})> \epsilon/4|\mathbb X_n]\Big ]d\theta\nn\\
&\leq& \sup_{\{g:D_H(g,f_{\theta_0})\leq\epsilon/4\}}\int_{D_H(f_{\theta},f_{\theta_0})>\delta^*}\frac{e^{-2nD_H(g_n,\theta)}\pi(\theta)}{\int e^{-2nD_H(g,\theta)}\pi(\theta)d\theta}d\theta\nn\\
&& +\Pi[D_H(g,f_{\theta_0})> \epsilon/4|\mathbb X_n].
\end{eqnarray}
Since $\Pi(\cdot|\mathbb X_n)$ is strongly consistent, the second term on the right hand side (RHS) of expression (\ref{eq:consistent1}) converges to $0$ almost surely.

Now  we show that the first term on the RHS of (\ref{eq:consistent1}) converges to $0$. It is sufficient to show that
\begin{equation}\label{eq:consistent2}
\frac
{\min_{\{g:D_H(g,f_{\theta_0})\leq\epsilon/4\}}
\int_{D_H(f_{\theta},f_{\theta_0})<\delta^*/4} e^{-2nD_H(g,f_{\theta})}\pi(\theta)d\theta}
{\max_{\{g:D_H(g,f_{\theta_0})\leq\epsilon/4\}}
\int_{D_H(f_{\theta},f_{\theta_0})>\delta^*} e^{-2nD_H(g,f_{\theta})}\pi(\theta)d\theta}\to \infty
\end{equation}
as $n$ goes to infinity. By triangular inequality, it is not hard to see that
$$
(\ref{eq:consistent2}) \leq C e^{2n\left[(\sqrt \epsilon /2-\sqrt {\delta^*})^2-(\sqrt \epsilon /2-\sqrt {\delta^*}/2)^2\right]},
$$
where $C=\int_{D_H(f_{\theta},f_{\theta_0})<\delta^*/4}\pi(\theta)d\theta
/\int_{D_H(f_{\theta},f_{\theta_0})>\delta^*}\pi(\theta)d\theta>0$ is some constant for any given $\delta$, and $\left[(\sqrt \epsilon /2-\sqrt {\delta^*})^2-(\sqrt \epsilon /2-\sqrt {\delta^*}/2)^2\right]>0$ since $\epsilon <\delta^*/4$.
$\Box$

\begin{remark}
The sufficient conditions under which the strong consistency holds for the Bayesian nonparametric model are given, eg. in \cite{ghosal99,wu10}.
\end{remark}

Now we show that the hierarchical model gives efficient estimator through the following theorem. We begin by stating some assumptions.
\begin{itemize}
\item[(C1)] $\{x:f_{\theta}(x)>0\}$ is the same for all $\theta\in \Theta$,

\item[(C2)] Interchanging the order of expectation with respect to $f_{\theta_0}$ and differentiation at $\theta_0$ are justified, so that
    $
    E_{\theta_0}\int \dot D_H(\theta_0,g)\Pi(g|\mathbb X_n)=0$, and   $ E_{\theta_0}\int \ddot D_H(\theta_0,g)\Pi(g|\mathbb X_n)=   - E_{\theta_0}(\int \dot D_H(\theta_0,g)\Pi(dg|\mathbb X_n))^2
    $
\item[(C3)]  $ I(\theta_0)\doteq E_{\theta_0}(\int \dot D_H(\theta_0,g)\Pi(dg|\mathbb X_n))^2>0.$

\item[(C4)]For any $\delta>0$, there exists $\epsilon>0$ such that the probability of the event
\begin{equation}
\sup \left\{\frac{1}{n}[l(\theta)-l(\theta_0)]:|\theta-\theta_0|\geq\delta\right\}\leq -\epsilon
\end{equation}
tends to 1 as $n\to \infty$, where $l(\theta)=\int D_H(g,\theta)d\Pi(g|\mathbb X_n)$.
\item[(C5)]The prior density $\pi$ of $\theta$ is continuous and positive for all $\theta\in \Theta$.

\item[(C6)] $E_\pi (|\theta|) < \infty$.
\end{itemize}

\begin{theorem}\label{thm:h}
If $\pi^*(t|\mathbb X)$ is the posterior density of $\sqrt n (\theta-\hat{\theta}_2)$ where
$\hat{\theta}_2$ is defined as in (\ref{eq:model1}) and the conditions of Theorems \ref{thm:1} to \ref{thm:3} hold,
\begin{itemize}
\item[{\rm(i)}] Then if (C1)-(C5) hold,
\begin{equation}\label{eq:88}
\int \left|\pi^*(t|\mathbb X)-\sqrt{I(\theta_0)}\phi\left[t\sqrt{I(\theta_0)}\right]\right|dt\stackrel{P}\to 0.
\end{equation}
\item[(ii)]If, in addition (C6) holds, then
\begin{equation} \label{eq:89}
\int \left(1+|t|\right)\left|\pi^*(t|\mathbb X)-\sqrt{I(\theta_0)}\phi\left[t\sqrt{I(\theta_0)}\right]\right|dt\stackrel{P}\to 0,
\end{equation}
where $\phi$ is the normal density.
\end{itemize}
\end{theorem}

{\sc Proof.}
We have that
\begin{eqnarray}
\lefteqn{\int \pi^*(t|\mathbb X)}\nn\\
&=&\int\frac
{\pi(\Hat{\theta}_2+\frac{t}{\sqrt n})\exp[-2n{\rm D_H}(\Hat{\theta}_2+\frac{t}{\sqrt n},g)]}
{\int\pi(\Hat{\theta}_2+\frac{s}{\sqrt n})\exp[-2n{\rm D_H}(\Hat{\theta}_2+\frac{s}{\sqrt n},g)]ds}
\frac{g(\mathbb X)\Pi(dg)} {\int g(\mathbb X)\Pi(dg)}dg\nn\\
&=&\int\frac
{\pi(\Hat{\theta}_2+\frac{t}{\sqrt n})\exp\big[-2n\big ({\rm D_H}(\Hat{\theta}_2+\frac{t}{\sqrt n},g)-{\rm D_H}(\Hat{\theta}_2,g)\big)\big]}
{\int\pi(\Hat{\theta}_2+\frac{s}{\sqrt n})\exp\big [-2n\big ({\rm D_H}(\Hat{\theta}_2+\frac{s}{\sqrt n},g)-{\rm D_H}(\Hat{\theta}_2,g)\big)\big]ds}\nn\\
&&\qquad\Pi(dg|\mathbb X_n).
\end{eqnarray}

We need to show
\begin{eqnarray}
 {\int}\left|\int\frac
{\pi(\Hat{\theta}_2+\frac{t}{\sqrt n})\exp\big[w(t)\big]}
{C_n}\Pi(dg|\mathbb X_n)
 -\sqrt{\frac{I(\theta_0)}{2\pi}}e^{\frac{-t^2I(\theta_0)}{2}}\right|dt
 \to 0&&
\end{eqnarray}
in probability, where $$C_n={\int\pi(\Hat{\theta}_2+\frac{s}{\sqrt n})\exp\big [w(t)\big]ds},$$ and
$$
w(t)=-2n\big [{\rm D_H}(\Hat{\theta}_2+\frac{t}{\sqrt n},g)-{\rm D_H}(\Hat{\theta}_2,g)\big].
$$
By the strong consistency, $\Pi(g|\mathbb X_n)\to \Ind _{f_{\theta_0}}$, where $\Ind_{f_{\theta_0}}$ is the indicator  function that is equal to $1$ while $g=f_{\theta_0}$ and $0$ otherwise. Therefore,
$$
\int\frac{\pi\left(\Hat{\theta}_2+\frac{t}{\sqrt n}\right)\exp\big[w(t)\big]}{C_n}\Pi(dg|\mathbb X_n)\to \frac{\pi(\Hat{\theta}_2+\frac{t}{\sqrt n})\exp\big[w_0(t)\big ]}{C_{n0}},
$$
where $w_0(t)$ and $C_{n0}$ correspond to $w(t)$ and $C_n$ for $g$ taking the value of $f_{\theta_0}$.

By Lemma \ref{lemma:J} (given in the appendix), we have that
\begin{equation}\label{eq:j1}
 \int\left|
{\pi\left(\Hat{\theta}_2+\frac{t}{\sqrt n}\right)\exp\big[w_0(t)\big]}-
{e^{\frac{-t^2I(\theta_0)}{2}}\pi(\theta_0)} \right|dt 
\to 0,
\end{equation}
and
\begin{equation}\label{eq:j2}
C_{n0} 
\to \pi(\theta_0)\sqrt{2\pi/I(\theta_0)}.
\end{equation}
By (\ref{eq:j2}), we have that
\begin{eqnarray}\label{eq:j3}
 \int\left|  \frac{e^{\frac{-t^2I(\theta_0)}{2}}\pi(\theta_0)}{C_{n0}}-\sqrt{\frac{I(\theta_0)}{2\pi}}
 e^{\frac{-t^2I(\theta_0)}{2}}
\right|dt 
\to 0.&&
\end{eqnarray}
By (\ref{eq:j1}) and (\ref{eq:j3}), the proof of part (i) is completed.

By the additional condition (C6), part (ii) holds from similar arguments.
$\Box$

Following this theorem, we demonstrate the asymptotic normality and efficiency of the expected {\em a posteriori} estimator for $\theta$ under this model when the data are generated from $f_{\theta_0}$. This result indicates that, asymptotically, the use of this hierarchical framework does not result in a loss of precision when the parametric model includes the true generating distribution.

\begin{theorem}\label{thm:6}
In addition to the assumptions of Theorem \ref{thm:h} assume that $\int | \theta | \pi(\theta)<\infty$. Let $\hat{\theta}_4=\int \theta \Pi^H(\theta|\mathbb X_n)d\theta$ be the Bayes estimate with respect to squared error loss. Then
\begin{itemize}
\item [(i)] $\sqrt n (\hat {\theta}_4-\hat{\theta}_2)\to 0$ in  probability,
\item [(ii)] $\sqrt n (\hat {\theta}_4-{\theta_0})$ converges in distribution to $N(0, 1/I(\theta_0))$.
\end{itemize}
\end{theorem}
{\sc Proof.} We have that
$$
\int (1+|t|)\left| \pi^*(t|\mathbb X_n)-\frac{\sqrt{I(\theta_0)}}{\sqrt{2\pi }}e^{\frac{-t^2I(\theta_0)}{2}}\right|dt
\to  0.
$$
Hence $\left|\int t\left| \pi^*(t|\mathbb X_n)-\frac{\sqrt{I(\theta_0)}}{\sqrt{2\pi }}e^{\frac{-t^2I(\theta_0)}{2}}\right|dt\right|
\to 0.$ Note that because
$$
\sqrt{\frac{I(\theta_0)}{2\pi}}\int te^{\frac{-t^2I(\theta_0)}{2}}dt=0
$$
we have $\int t \pi^*(t|\mathbb X_n)\to 0$.
Note that
$$
\hat{\theta}_4=\int \theta \Pi^H(\theta|\mathbb X_n)d\theta=\int (\hat{\theta}_2+t/\sqrt n)\pi^*(t|\mathbb X_n)dt
$$
and hence $\sqrt n (\hat{\theta}_4-\hat {\theta}_2)=\int t \pi^*(t|\mathbb X_n)dt$.
Assertion (ii) follows (i) and the asymptotic normality of $\hat {\theta}_2$ discussed earlier.
$\Box$

\section{Robustness properties}

In this section, we examine the robustness properties of the proposed hierarchical model. In this, we will follow the notion of ``outlier rejection'' from the Bayesian analysis of robustness, but note its similarity to frequentist propositions.  In frequentist analysis, robustness is usually measured by the influence function and breakdown point of estimators.  These have been used to study robustness in minimum Hellinger distance estimators in \cite{Beran77} and in more general minimum disparity estimators in \cite{ParkBasu04} and \cite{Hooker11}.

In Bayesian inference, robustness is labeled ``outlier rejection'' and is studied under the framework of ``theory of conflict resolution''. \cite{deFinetti61} described how outlier rejection could take place quite naturally in Bayesian context. He did not demonstrate formally that such behavior would happen, but described how the posterior distribution would  be influenced less and less by more and more distant outlying observations. Eventually, as the separation between the outliers and the remainder of the observations approached infinity, their influence on the posterior distribution would become negligible, which is a rejection of the outliers. \cite{dawid73} gave conditions on the model distribution and the prior distribution which ensure that the posterior expectation of a given function tends to its prior expectation. Note that such ignorability of extreme outliers is regardless of prior information. \cite{Ohagan79} generalized the \cite{dawid73}'s work, and introduced the concept of outlier-proneness. His result can be easily extended to a more general cases, where the observed data is considered as several subgroups, by applying the concept ``credence'' introduced by \cite{Ohagan90}. While \cite{Ohagan90}'s results are only about symmetric distribution, \cite{desgagne07} gave corresponding results covering a wider class of distributions with tails in the general exponential power family. These results provided  a complete theory for the case of many observations and a single location parameter. Unfortunately, there are only limited results for Bayesian hierarchical models. \cite{angers91} proved that outlier rejection occurs in some particular hierarchical models when model for the mid level random variables is Cauchy distribution and the model for the observations given mid level random variables is normal. \cite{choy97} gave numerical examples of the same behavior when Cauchy distribution replaced by some other heavy-tailed distributions.

For the hierarchical Hellinger model {considered in this paper}, consider $m$ groups of observations, identified by disjoint subsets $S_j,j=1,2,\ldots, m$, of the indices. Thus, $\cup_{j=1}^mS_j=\{1,2,\ldots,n\}$ and $S_j \cap S_{j'}=\varnothing$ when $j\neq j'$. Let $n_{S_i}$ denote the number of observations belong to $S_i$. We suppose that the observations in group 1 remain fixed while the other groups move increasingly far apart from the first group and from each other. Formally, for $i\in S_j$ we write $x_i=\tilde x_j+z_i$, so that $\tilde x_j$ of a reference point for group $j$ and the $z_i$'s denote deviations of the observations from their respective reference point. Then we let the reference points $\tilde x_2,\tilde x_3,\ldots, \tilde x_m$, tend to $\infty$ and/or $-\infty$ such that the separations $|\tilde x_j-\tilde x_{j'}|$ all tend to infinity, while the $z_i$'s remain fixed. {Our limiting results in this section are all under this scenario.} We now demonstrate that the posterior distribution of $\theta$ tends to the posterior defined as
\begin{eqnarray}\label{eq:robusttarget}
\lefteqn{\quad\Pi^H(\theta|\mathbb X_{S_{1b}})}\\
 &=&\iint_g     \frac{\pi(\theta)e^{-2n\sqrt b D_H(g,\theta)}}{\int\pi(\theta)e^{-2n\sqrt b D_H(g,\theta)}d\theta}\frac{\Pi(d g)g(\mathbb X_{S_1})}{\int\Pi(d g)g(\mathbb X_{S_1})}dgd{\rm B}(b;n_1+1,n-n_1),\nn
\end{eqnarray}
that would arise given only the information sources in group 1, and where ${\rm B}(b;n_1+1,n_1-1)$ denotes a Beta distribution with parameters $(n_1+1,n-n_1)$.

\begin{theorem}\label{thmrob1}
Suppose that the hierarchical Hellinger  Bayesian model  is defined as
(\ref{eq:hp}), and the prior for density estimation is specified to be
a  Dirichlet mixture, with kernel density function $K(x,\tau)$
and Dirichlet process prior $D_{\alpha}$, where $\alpha= M \bar
{\alpha}$, $M$ is a positive scale constant and $\bar{\alpha}$ is a probability measure, then $\Pi^H(\theta|\mathbb X_n)\to \Pi^H(\theta|\mathbb X_{S_{1b}})$ in probability as the reference points tend to $\pm \infty$.
\end{theorem}
{\sc Proof.}   The Dirichlet mixture prior for density estimation {models} the observation $\{x_1,\ldots,x_n\} $ {as} i.i.d.  follow a density function  $g(x) = \int K(x,\tau)dP(\tau)$ for given $P(\tau)$, where $P$, the mixing distribution, is given prior distribution $\mathscr D_{\alpha}$.

We can express the posterior distribution for mixing distribution $P$ as
$$
\Pi(P|\mathbb X_n)=\int (\mathscr
D_{\alpha+\sum_{i=1}^n\Ind_{\tau_i}})H(d\underline{\tau}|\mathbb X_n),
$$
where $H$ denotes the posterior distribution  for hidden variable
$\tau$. Let $C_1,\ldots,C_{N(\mathbb P)}$ be a partition of $\{1,2,\ldots,n\}$ and $e_j$'s be the number of the elements in $C_j$'s such that for any $i \neq i^*$ and $\tau_i=\tau_{i^*}$, we have $\tau_i\in C_j$ and $\tau_{i^*}\in C_j$ for some $j$. That is, the $C_j$ correspond to the indices of $\tau_i$ that are repeated.  The density of $H(\underline{\tau}|\mathbb X_n)$ is
\begin{equation}
h(\underline{\tau}|\mathbb
X_n)=\frac{\prod_1^{N(\mathbb P)}\left(\alpha(\tau_j)(e_j-1)!\prod_{l\in
C_j}K(x_l,\tau_j)\right)}{\sum_{\mathbb P}
\int\prod_1^{N(\mathbb P)}\alpha(\tau_j)(e_j-1)!K(x_l,\tau_j)d\tau_j},
\end{equation}
where $\underline {\tau}=\{\tau_1,\cdots, \tau_n\}$. For any $\epsilon>0$ there exists $C>0$ such that
\begin{equation}\label{eq:27}
\int_{\|\underline{\tau}-\mathbb X_n\|>C} H(\underline {\tau}|\mathbb X_n)d\underline {\tau}<\epsilon.
\end{equation}
Hence, as $\tilde x_2, \ldots,\tilde x_m$ go to $\pm \infty$ and $|\tilde x_j-\tilde x_{j'}|\to \infty$ when $j\neq j'$,
\begin{equation}\label{eq:28}
\int \left |\prod_{i=1}^m \frac{\prod_1^{N(\mathbb P_{S_i})}\left(\alpha(\tau_j)(e_j-1)!\prod_{l\in
C_j}K(x_l,\tau_j)\right)}{\sum_{\mathbb P_{S_i}}
\int\prod_1^{N(\mathbb P_{S_i})}\alpha(\tau_j)(e_j-1)!K(x_l,\tau_j)d\tau_j} - h(\underline {\tau}|\mathbb X_n)\right |d \underline {\tau}\to 0
\end{equation}
in probability, where $\mathbb P_{S_i}$ denotes the partition on $\{1,2,\ldots, n_{S_i}\}$ {and $N(\mathbb P_{S_i})$ is the number of subsets of the partition and the result occurs since $K(x_l,\tau_j) \to 0$ for all except one partition for any $j$.}
Here, the first term represents the evaluation of densities calculated on each subgroup separately. As the subgroups become more separated, this approximates the calculation on the entire data set.

The Dirichlet process mixture prior and the corresponding posterior  assign probability measure on the random probability density function via the mapping: $P\mapsto \int K(\cdot)dP$, such that the distribution of the mixing distribution $P$  induces the distribution of the random probability density function of the observations.  Therefore, the Hellinger posterior  distribution is
\begin{eqnarray*}
\lefteqn{\Pi^H(\theta|\mathbb
X_n)}\\
&=&\int \frac{\pi(\theta)e^{-2nD_H(g,\theta)}}{\int\pi(\theta)e^{-2nD_H(g,\theta)}d\theta}\frac{\Pi(
g)g(\mathbb X_n)}{\int\Pi(d g)g(\mathbb X_n)}dg\\
&=&\int\frac{\pi(\theta)e^{-2nD_H(\int
K(t,\xi)dP(\xi),f_{\theta}(t))}}{\int\pi(\theta)e^{-2nD_H(\int
K(t,\xi)dP(\xi),f_{\theta}(t))}d\theta}d  \left[\int(\mathscr
D_{\alpha+\sum_{i=1}^n\Ind_{\tau_i}})H(d\underline{\tau}|\mathbb X_n) \right]\\
&=&\int \left[ \int \frac{\pi(\theta)e^{-2nD_H(\int
K(t,\xi)dP(\xi),f_{\theta}(t))}}{\int\pi(\theta)e^{-2nD_H(\int
K(t,\xi)dP(\xi),f_{\theta}(t))}d\theta}d(\mathscr
D_{\alpha+\sum_{i=1}^n\Ind_{\tau_i}}) \right]H(d\underline{\tau}|\mathbb X_n),
\end{eqnarray*}
and the last line is by Fubini's theorem {since $D_H$ is bounded below and $\pi(\theta)$ is bounded above}.

Let $\mathscr B_k=\{B_{k1},\ldots,B_{kk}\}$ be a sequence of partitions of $\mathscr B=\cup_{i=1}^n [X_i-C,X_i+C]$, $v_k=\max\{{\rm volume}(B_{kj})|1\leq j\leq k\}$, and $\lim_{k\to \infty}v_k=0$. By the definition of the Dirichlet process we write
\begin{eqnarray}\label{eq:bct}
\lefteqn{\quad\Pi^H(\theta|\mathbb
X_n)}\\
&=&\int\left [\lim_{k\to\infty}\int \frac{\pi(\theta)e^{-2nD_H(\sum
K(t,\xi_{ki})P(B_{ki}),f_{\theta}(t))}}{\int\pi(\theta)e^{-2nD_H(\sum
K(t,\xi_{ki})P(B_{ki}),f_{\theta}(t))}d\theta}d(\mathbb
D_k)\right ]H(d\underline{\tau}|\mathbb X_n), \nn
\end{eqnarray}
where $\mathbb D_k$ denotes a Dirichlet distribution with parameter
\[
\left(\left[\alpha+\sum_{i=1}^n\Ind_{\tau_i}\right](B_{k1}),\ldots,\left[\alpha+\sum_{i=1}^n\Ind_{\tau_i}\right](B_{kk})\right),
\]
where $\xi_{ki}\in B_{ki}$ and $(P(B_{k1}),\ldots,P(B_{kk}))\sim \mathbb D_k$.
By the bounded convergence theorem, \eqref{eq:bct} becomes
\begin{eqnarray}\label{eq:291}
\lefteqn{\quad\Pi^H(\theta|\mathbb
X_n)}\\
&=&\lim_{k\to\infty}\int \left[\int \frac{\pi(\theta)e^{-2nD_H(\sum
K(t,\xi_{ki})P(B_{ki}),f_{\theta}(t))}}{\int\pi(\theta)e^{-2nD_H(\sum
K(t,\xi_{ki})P(B_{ki}),f_{\theta}(t))}d\theta}d(\mathbb
D_k)\right ]H(d\underline{\tau}|\mathbb X_n).\nn
\end{eqnarray}
By (\ref{eq:27}), considering the case that  $\tau_i\in [x_i-C,x_i+C]$ for every $i$, we have that
\begin{eqnarray}\label{eq:30}
\lefteqn{\int \frac{\pi(\theta)e^{-2nD_H(\sum_{i=1}^k
K(t,\xi_{ki})P(B_{ki}),f_{\theta}(t))}}{\int\pi(\theta)e^{-2nD_H(\sum_{i=1}^k
K(t,\xi_{ki})P(B_{ki}),f_{\theta}(t))}d\theta}d(\mathbb
D_k)}\\
&=& \int \frac{\pi(\theta)e^{-2nD_H(\sum_{i=1}^k
K(t,\xi_{ki})P(B_{ki})\Ind_{\xi_{ki}\in S_1},f_{\theta}(t))}}{\int\pi(\theta)e^{-2nD_H(\sum_{i=1}^k
K(t,\xi_{ki})P(B_{ki})\Ind_{\xi_{ki}\in S_1},f_{\theta}(t))}d\theta}d(\mathbb
D_k).\nn
\end{eqnarray}
This is the (approximated) hierarchical Hellinger posterior using only the observations from $S_1$.
To see this, let
$$
a(t)=\sum_{i=1}^k K(t,\xi_{ki})P(B_{ki})\Ind_{\xi_{ki}\in S_1},
$$  and
$$
r(t)=\sum_{i=1}^k K(t,\xi_{ki})P(B_{ki})\Ind_{\xi_{ki}\notin S_1},
$$
and note that $a(t)+r(t)=\sum_{i=1}^k K(t,\xi_ki)P(B_ki)$.
Then we have that
$$
D_H\left([a(t)+r(t)],f_{\theta}(t)\right)
=
2-2\int [a(t)+r(t)]^{1/2}f^{1/2}_{\theta}(t)dt.
$$
As $\tilde x_i$'s, $i=2,\ldots,m$, go to infinity, the $B_{ki}$ become further separated and $\min\{a(t),r(t)\}\to 0$. Hence,
$$[a(t)+r(t)]^{1/2}\to a^{1/2}(t)+b^{1/2}(t).$$
For the same reason, $\int b^{1/2}(t)f^{1/2}_{\theta}(t)dt=0$. Therefore,
$$
D_H([a(t)+r(t)], f_{\theta}(t))=2-2\int a(t)^{1/2}f^{1/2}_{\theta}(t)dt.
$$
Let $b=\int a(t)dt$, then $D_H([a(t)+r(t)], f_{\theta}(t))=D_H(a(t), f_{\theta}(t))+1-b$.
Noting that $1-b$ is cancelled, we have (\ref{eq:30}). Rewrite (\ref{eq:30}) as
\begin{equation}\label{eq:31}
\int \frac{\pi(\theta)e^{-2n\int(\sum_{i=1}^k
K(t,\xi_{ki})P(B_{ki})\Ind_{\xi_{ki}\in S_1}f_{\theta}(t))dt}}{\int\pi(\theta)e^{-2n\int(\sum_{i=1}^k
K(t,\xi_{ki})P(B_{ki})\Ind_{\xi_{ki}\in S_1}f_{\theta}(t))dt}d\theta} d(\mathbb
D_k).
\end{equation}
Assuming that we have only the observations in $S_1$, the Hellinger posterior should be
\begin{eqnarray*}
\lefteqn{\Pi^H(\theta|\mathbb
X_{S_1})}\\
&=&\lim_{k\to\infty}\int \left[\int \frac{\pi(\theta)e^{-2nD_H(\sum
K(t,\xi_{ki})P(B_{ki}),f_{\theta}(t))}}{\int\pi(\theta)e^{-2nD_H(\sum
K(t,\xi_{ki})P(B_{ki}),f_{\theta}(t))}d\theta}d(\mathbb
D_k)\right ]H(d\underline{\tau}|\mathbb X_{S_1}),
\end{eqnarray*}
where  $B_{ki}, \xi_{ki}$, and $ \mathbb D_k$ are defined as before for the observation $\mathbb X_{S_1}$
and $H(d\underline{\tau}|\mathbb X_{S_1})$ is defined as  (\ref{eq:28}) with $m=1$.
Based on the aggregation property of Dirichlet distribution and writing the Hellinger distance in its integral form, we see that (\ref{eq:31}) is equal to
\begin{equation}\label{eq:32}
\int \frac{\pi(\theta)e^{-2n\sqrt b\int(\sqrt{\sum_{i=1}^k
K(t,\xi_{ki})P(B_{ki})} - \sqrt{f_{\theta}(t)})^2dt}}{\int\pi(\theta)e^{-2n\sqrt b\int\sqrt{(\sum_{i=1}^k
K(t,\xi_{ki})P(B_{ki})}-\sqrt{f_{\theta}(t)})^2dt}d\theta}d {\rm Beta}(b) d(\mathbb
D_k),
\end{equation}
where the parameters of the beta distribution are $(n_1+1,n-n_1)$ and $\mathbb D_k$ denotes the Dirchlet distribution on $\mathscr B$ defined on $\mathbb X_{S_1}$.
Now, substitute (\ref{eq:32}) into (\ref{eq:291})  and note that $h(\underline {\tau}|\mathbb X_n)=h(\underline {\tau}_{S_1}|\mathbb X_{S_1})\times h(\underline {\tau}_{-S_1})$, where $h(\tau_{-S_1})$ denotes the density function on the component of $\tau$ corresponding to all observations that are not in $S_1$. Since the Hellinger distance is bounded from above, the Fubini's Theorem applies and the proof is completed.
$\Box$

\begin{remark}
When we only observe $n_1$ uncontaminated observation, the Hellinger posterior distribution is
$$
\int_g     \frac{\pi(\theta)e^{-2n_1D_H(g,\theta)}}{\int\pi(\theta)e^{-2n_1D_H(g,\theta)}d\theta}\frac{\Pi(d g)g(\mathbb X_{S_1})}{\int\Pi(d g)g(\mathbb X_{S_1})}dg.
$$
When we have $n$ observations with $n_1$ of them uncontaminated, the Hellinger posterior distribution tends to (25), where the expectation of $b$ is $\frac{n_1+1}{n+1}$. This is approximately the distribution that would result from simply ignoring the outliers.
\end{remark}

\section{Sampling from Posterior Distribution on Density Functions}
The Dirichlet process Gaussian probability density kernel mixture prior is one of the most used prior for probability  density  function estimation in practice. The computation of the posterior is not trivial and there is quite large amount of researches on this topic.
We use an algorithm based on the stick breaking process {and} WinBugs to carry out the MCMC sampling for the posterior of the mixing distribution. See, Blei and Jordan (2006), Ishwaran and James (2001), Kemp (2006) and Xu (2006) for more details of stick breaking algorithm for Dirichlet Mixture prior.

In this paper, we use the following prior for Bayesian nonparametric density estimation and the proposed hierarchical Hellinger methods.
\begin{eqnarray}
x_i|\mu_i,\Sigma_i&\sim& N(\mu_i,\Sigma_i) i=1,\ldots,n,\\
(\mu_i,\Sigma_i)|g&\sim&G,\nn\\
G|\alpha,G_0&\sim&DP(\alpha,G_0),\nn
\end{eqnarray}
where $G_0=N(\mu|m_1,\kappa_0^{-1}\Sigma)InvWhishart(\Sigma|\nu_1,\psi_1)$, $\kappa_0\sim Gamma(0.5,50)$, $m_1=0$, $\alpha=0$, $\psi_1=2$ and $\nu_1=4$.

Once a sample of $g$ from the posterior distribution $\Pi(g|\mathbb{X})$ is obtained, we use Metropolis algorithm to draw samples of $\theta$ from $\frac{\pi(\theta)e^{-2nD_H(\theta,g)}}{\int \pi(\theta)e^{-2nD_H(\theta,g)}d\theta}$ for each $g$ in the previous obtained sample of $g$. The squared Hellinger distance $D_H(\cdot,\cdot)$ is calculated by standard numerical quadrature method.

For any given data $\mathbb X_n$, by collecting all the samples of $\theta$ obtained as described above, we have a sample of $\theta$ follows the hierarchical Hellinger posterior distribution. We report the arithmetic average of the sample as the hierarchical Hellinger estimate of the parameter and the $2.5\%$ and $97.5\%$ quartiles of the sample as the $95\%$ credible interval of the estimate.

\section{Simulations and Real Data}





To examine the computational feasibility and finite sample size behavior of $\hat{\theta}$, several numerical experiments were carried out. The Dirichlet mixture prior as described in the previous section was used in all these experiments to conduct the posterior  distribution of the probability density functions, from which samples were drawn. We undertook a simulation study for i.i.d. data from Gaussian distribution. 1000 sample data sets of size 20 from a $N(5,1)$ population were generated. For each sample data set, a Metropolis algorith was run for 2,000,000 steps using a $N(0,.5)$ proposal distribution and a $N(0,25)$ prior, placing the true mean on prior standard deviation above the prior mean. Expected a posteriori estimates for the sample mean were obtained along with $95\%$ credible intervals from every sample in the second half of the MCMC chain. Outlier contamination was investigated by reducing the last one, two or five elements in the data set by 3,5 or 10. This choice was made so that both outliers and prior influence the Hierarchical Hellinger method in the same direction. The analytic posterior without the outliers is normal with mean 4.99 (equivalently, bias of -0.01) and standard deviation 0.223.

The results of this simulation are summarized in Tables 1 (uncontaminated data) and 2 (contaminated data).

\begin{table}[ht]
\caption{A simulation study for a normal mean using the usual posterior and the Hierachical Hellinger posterior. Columns give the bias and variance of the posterior mean, coverage and the central $95\%$ credible interval based on 1000 simulations. }
\centering
\begin{tabular}{c c c c c }
\hline\hline
 & Bias & SD &Coverage &Length  \\
\hline
Posterior & -0.015 & 0.222 & 0.956 &0.873 \\
Hellinger & -0.015 & 0.227 & 0.955 &0.937 \\
\hline
\end{tabular}
\end{table}

{
 \begin{table}[ht]
\caption{Results for contaminating the data sets used in Table 1 with outliers. 1, 2, and 5 outliers (large columns) are added at locations -3, -5 and -10 (column Loc) for the posterior and Hierachical Hellinger posterior.}
\centering
\tiny{
\begin{tabular}{c c | c c c| c c c| c c c  }
\hline\hline
\# of Outliers&&&1&&&2&&&5&\\
& & Bias & SD &Cov  & Bias & SD &Cov & Bias & SD &Cov \\
\hline
Posterior &-3& -0.147 & 0.219 & 0.883& -0.301 & 0.206 & 0.722&-0.636 & 0.182 & 0.100 \\
          &-5& -0.248 & 0.219 & 0.778& -0.492 & 0.206 & 0.375&-1.054 & 0.182 & 0.001 \\
          &-10& -0.519 & 0.219 & 0.360& -0.965 & 0.207 & 0.004&-2.092 & 0.182 & 0.000 \\
\hline\\
Hellinger &-3& -0.108 & 0.249 & 0.922& -0.197 & 0.278 & 0.852 & 0.239 &0.303& 0.771 \\
          &-5& -0.027 & 0.240 & 0.940& -0.040 & 0.257 & 0.923 & 0.024 &0.307& 0.856 \\
          &-10& -0.014 & 0.234 & 0.948& -0.019 & 0.249 & 0.937& 0.018 &0.287& 0.886 \\
\hline


\hline

\end{tabular}
}
\end{table}

}

 We now apply the hierarchical method to a real data example. The data come from one equine farm participating in a parasite control study in Denmark in 2008.  Eggs of equine Strongyle parasites in feces were counted before and after the treatment with the drug Pyrantol. The full study is presented in \cite{nielsen10}. The raw data is given in the Table \ref{data}. For each horse, we can calculate the observed survival rate of the parasite. We assume that the observed log odds follows a normal distribution and  estimate the mean and variance of the normal distribution by the hierarchical   Hellinger method.
\begin{table}[ht]
  \begin{center}
 \begin{tabular}{|c|ccccccc|}
  \hline
  Horse&1&2&3&4&5&6&7\\
  \hline
  Before treatment&2440&1000&1900&1820&3260&300&660\\
  After treatment&580&320&400&160&60&40&120\\
  \hline
\end{tabular}
\end{center}
  \caption{The first row of the table contains the fecal egg counts for seven horses on one equine farm before they took drug Pyrantol. The second row of the table contains the fecal egg counts for these horses after they took drug Pyrantol.}
  \label{data}
\end{table}

\begin{figure}[ht]
\begin{center}
   \includegraphics[width=0.49\linewidth]{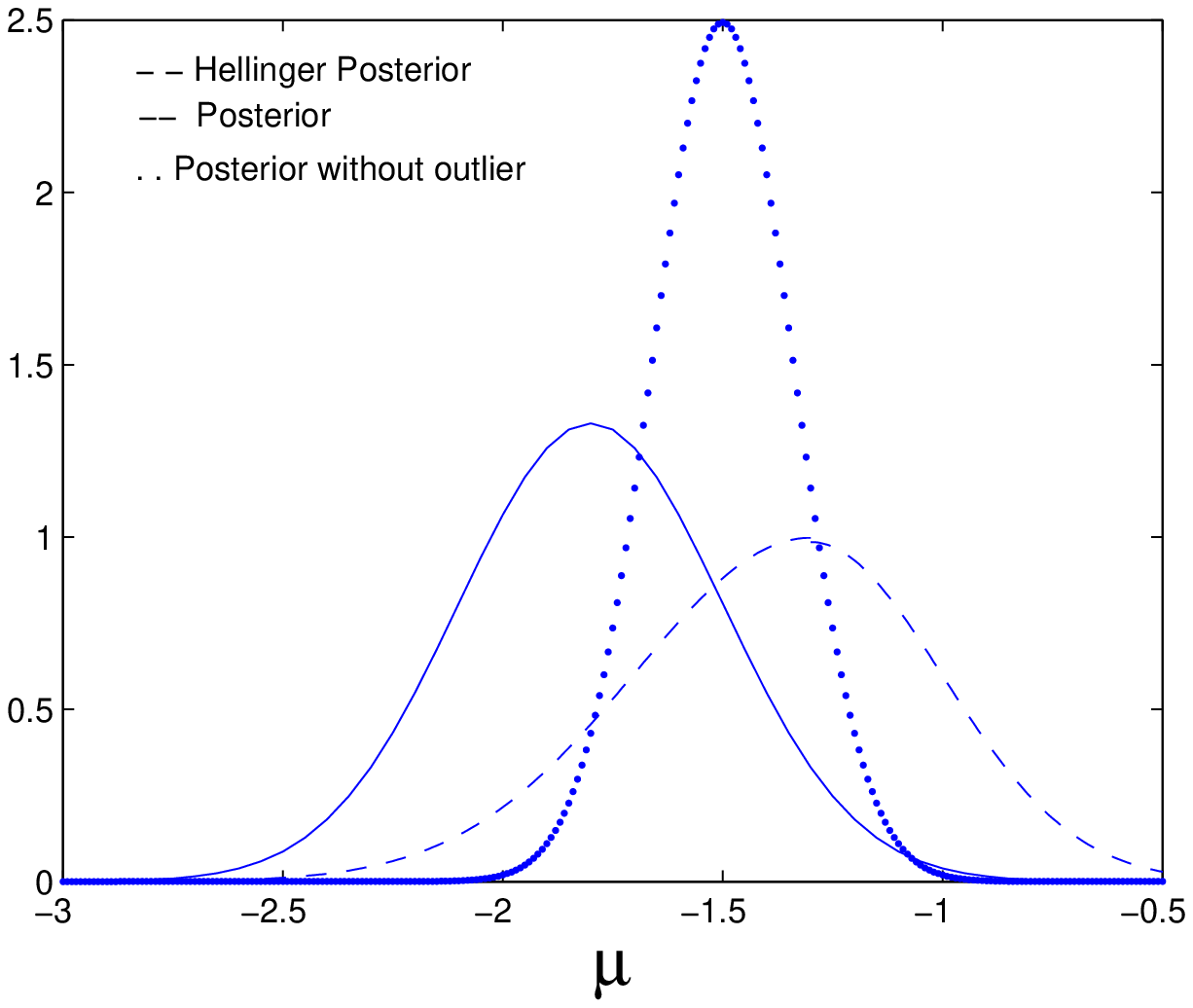}
   \includegraphics[width=0.49\linewidth]{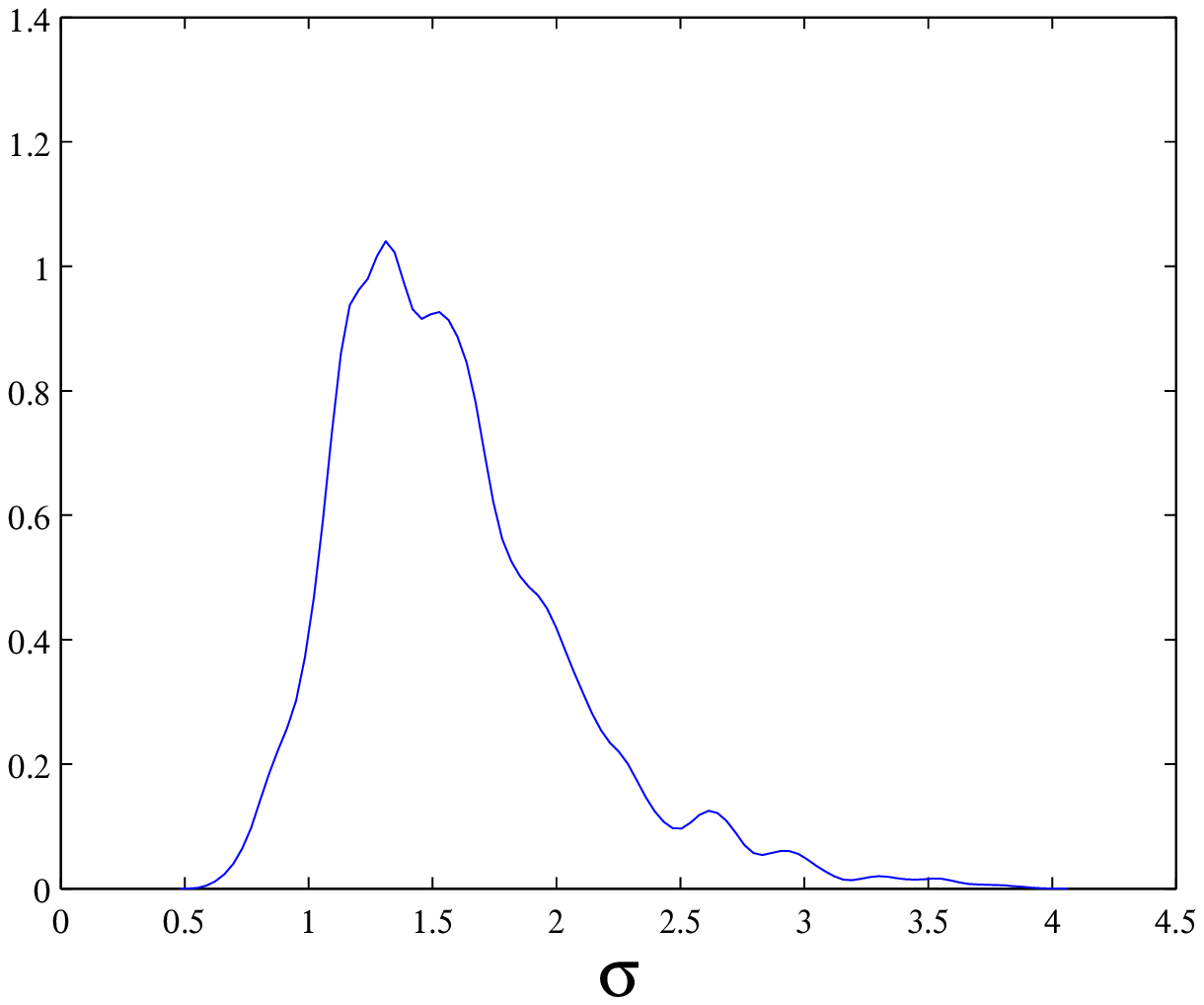}
\end{center}
\caption{Posterior distributions for the parasite data. Left: posteriors for $\mu$ with and without an outlier and the Hellinger posterior. Right: posteriors for $\sigma$. }\label{fig:2}
\end{figure}

To complete the hierarchical  Hellinger Bayesian estimation, we assign $\mu$ a prior $N(0,5)$ and $\sigma^2$ a prior $Gamma(3,0.5)$, where $\mu$ and $\sigma$ denote the parameter of the normal distribution, which the observed log odds are assumed to follow. We plotted the Hellinger posterior distribution of these two values  in Figure \ref{fig:2}.

\begin{figure}[ht]
\begin{center}
   \includegraphics[width=0.39\linewidth]{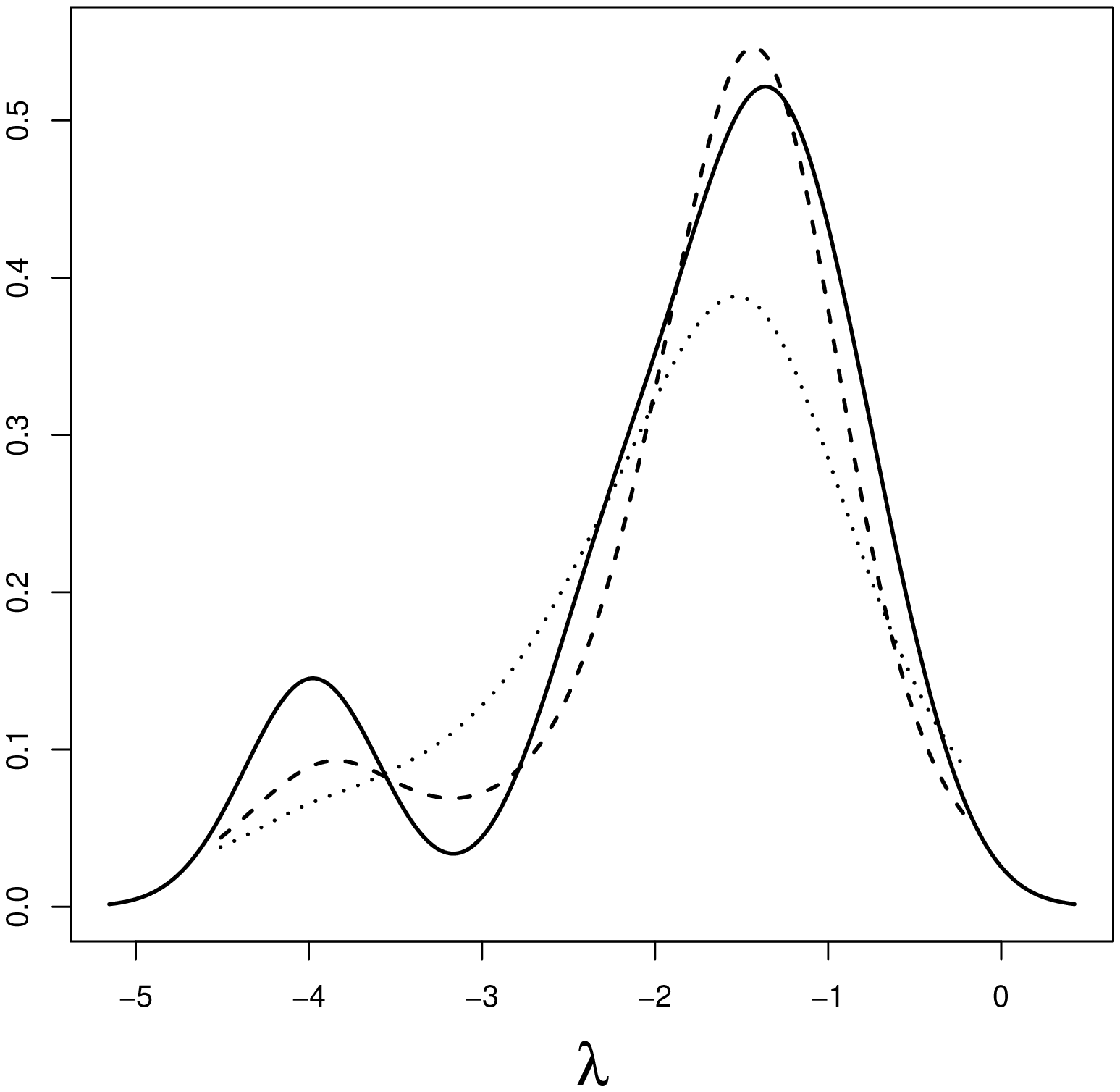}
   \includegraphics[width=0.49\linewidth]{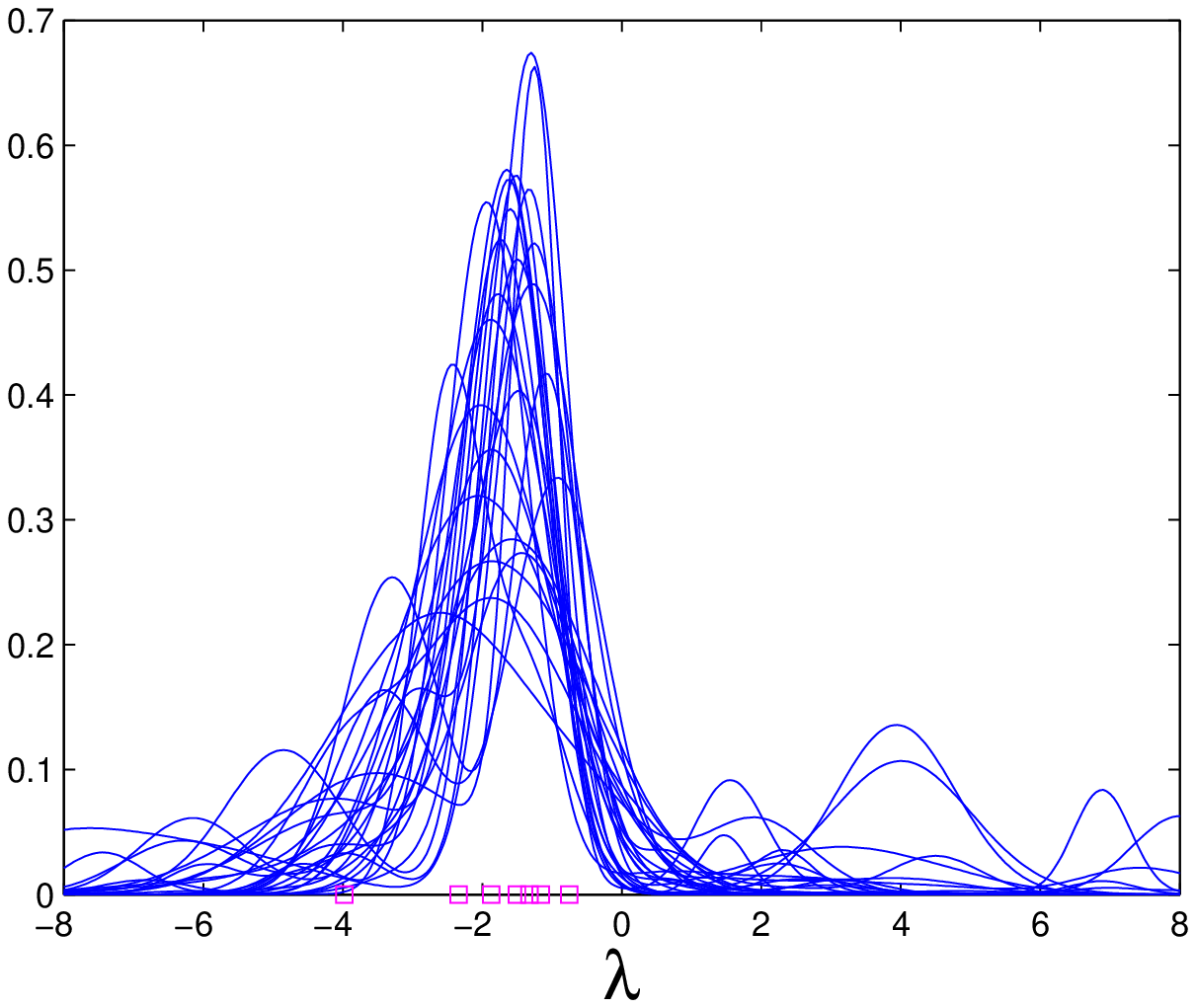}
\end{center}
\caption{Left:  Nonparametric density estimations. The smooth line is based on the classical kernel density estimation with default bandwidth chosen by R. The dashed and dotted lines represented density function estimations based on Dirichlet process mixture prior with different flexibility. To show that Hierarchical Hellinger method is not rely on the bimodal property of the density, we choose the prior setting related to the single modal density estimation for our real data study.  Right: samples of $g$ based on draws from the posterior updated from Dirichlet process normal mixture prior, for which the density function estimation is single modal, demonstrating an outlier at $-4$ and much more flexible than classical kernel density estimates.}\label{fig:1}
\end{figure}

With classical method, the mean of this data is -1.85 with standard deviation 1.07. If we removed the suspected outlier, -3.98, the mean and the standard deviation are -1.49 and 0.56.
If we assume that the data without the suspected outlier gives the value around the true standard deviation, then the estimate we obtained based on the nonparametric prior with standard normal distribution as base measure underestimates it while the other overestimates it, with corresponding values about 0.70 and 0.47 after re-exponentiating $\log \sigma$.

\section{Discussion}

In this paper we argue that the hierarchical framework described here represents the natural means of combining the robustness properties of minim disparity estimates with Bayesian inference. In particular, by modifying the Hellinger likelihood methods of \cite{Hooker11} to incorporate a Bayesian non-parametric density, we are able to obtain a complete Bayesian model.  Furthermore, we demonstrate that this model retains the desirable properties of minimum-disparity estimation: it is robust to outlying observations but retains the precision of a parametric estimator when such observations are not present.  Indeed, this framework represents a more general means of combining parametric and non-parametric models in Bayesian analysis: the parametric model representing an approximation to the truth that informs, but does not dictate, a non-parametric representation of the data.

Despite its advantages, there remains considerable future problems to be addressed. While we have restricted our attention to Hellinger distance for the sake of mathematical convenience, the same arguments can be extended to general disparities. These take the form
\[
D(g,f_{\theta}) = \int G\left( \frac{g(x)}{f_\theta(x)} -1 \right) f_\theta(x) dx \leq G(0) + C \int |g(x) - f_{\theta}(x) |dx
\]
for appropriate $G$. Using this $L_1$ bound, Lemma \ref{lem:1} can be generalized so that the estimate incorporating a random histogram posterior is closer than $o(n^{-1})$ to a minimum disparity estimator with a kernel density estimate. Since the latter is efficient we can apply Theorem \ref{thm:2}. Calculations for the hierarchical Hellinger model can also be generalized to this context in a manner similar to \cite{Hooker11}. More general disparities are of interest in some instances. Hellinger distance corresponds to $G(x) = (\sqrt{x+1} -1)^2$ which while insensitive to large values of $g(x)/f_\theta(x)$ is senstive to small values, also called ``inliers''. By contrast, the negative exponential disparity corresponds to $G(x) = e^{-x}$ is insensitive to both.

In other theoretical  problems, we have so far employed random histogram priors or Dirichlet process mixture priors as convenient (for efficiency and robustness respectively); general conditions on Bayesian nonparametric densities under which these results hold must be developed. See \cite{castillo13} for recent developments. Computationally, we have applied a simple rejection sampler to non-parametric densities generated without reference to the parametric family; improving the computational efficiency of our estimators will be an important task.  Methodologically, \cite{Hooker11} demonstrate the use of minimum Disparity methods and Bayesian inference on regression and random-effects models -- considerably expanding the applicability disparity methods. These methods depended on either nonparametric estimates of conditional densities, or kernel densities over data transformations that dependent on parameters. Finding computationally feasible means of employing non-parametric Bayesian density estimation in this context is expected to be a challenging and important problem.

\bibliography{mybib}

\bibliographystyle{abbrvnat}

\section*{Appendix}

\begin{lemma}\label{lemma:J}
If conditions (B1) and (C1) -(C5) hold, then we have
\begin{equation}\label{eq:j12}
 \int\left|
{\pi\left(\Hat{\theta}_2+\frac{t}{\sqrt n}\right)\exp\big[w_0(t)\big]}-
{e^{\frac{-t^2I(\theta_0)}{2}}\pi(\theta_0)} \right|dt 
\to 0,
\end{equation}
\end{lemma}

{\sc Proof}. Let $h_n=2 \int \ddot D_H (\hat {\theta}_3,g)\Pi(dg|\mathbb X_n)$. First, we show that as $n\to \infty$, $h_n\to I(\theta_0)$. We see that
\begin{eqnarray}
h_n&=&\iint\left( f_{\hat{\theta}_2}^{-3/2}(t)f^{\prime2}_{\hat{\theta}_2}(t) -2f_{\hat{\theta}_2}^{-1/2}(t)f^{\prime \prime2}_{\hat{\theta}_2}(t)\right )g^{1/2}(t)dt\Pi(dg|\mathbb X_n)\nn\\
&=&\int f_{\hat{\theta}_2}^{-3/2}(t)f^{\prime2}_{\hat{\theta}_2}(t)\int g^{1/2}(t)\Pi(dg|\mathbb X_n) dt \nn\\
&& -2\int f_{\hat{\theta}_2}^{-1/2}(t)f^{\prime \prime2}_{\hat{\theta}_2}(t) \int g^{1/2}(t)\Pi(dg|\mathbb X_n)dt.
\end{eqnarray}
By assumption $\Pi(dg|\mathbb X_n)\to \Ind _{g_0}$, which implies that
$$
\int g^{1/2}(t)\Pi(dg|\mathbb X_n)dt\to g_0^{1/2}(t).
$$
By the continuity of $f$ on $\theta$, and $\hat{\theta}_2\to \theta_0$ in  probability. We have that $f_{\hat{\theta}_2}^{-3/2}(t)\to f_{{\theta}_0}^{-3/2}(t)$,
$f_{\hat{\theta}_2}^{-1/2}(t)\to f_{{\theta}_0}^{-1/2}(t)$, $f^{\prime \prime2}_{\hat{\theta}_2}(t)\to f^{\prime \prime2}_{{\theta}_0}(t)$ and $f^{\prime 2}_{\hat{\theta}_2}(t)\to f^{\prime 2}_{{\theta}_0}(t)$ in $P_{f_{\theta_0}}$ probability. Therefore, we have that
$$
h_n\to \int f_{{\theta}_0}^{-1}(t)f^{\prime2}_{{\theta}_0}(t) dt-2\int f^{\prime \prime2}_{{\theta}_0}(t)dt=I(\theta_0),
$$
in  probability.

We see that to verify (\ref{eq:j12}) it is sufficient to show that
\begin{equation}\label{eq:tar}
\int\left|
{\int \pi\left(\Hat{\theta}_2+\frac{t}{\sqrt n}\right)\exp\big[w(t)\big]}\Pi(g|\mathbb X_n)-
{e^{\frac{-t^2h_n}{2}}\pi(\hat{\theta}_2)} \right|dt 
\to 0.
\end{equation}

To show (\ref{eq:tar}), given $\delta,c>0$ we break $\mathbb R$ in to three regions:
\begin{itemize}
\item [] $A_1=\{t:|t|<c\log \sqrt n\},$
\item [] $ A_2=\{t: c\log \sqrt n<|t|<\delta\sqrt n\}$, and
\item [] $ A_3=\{t:t:|t|>\delta\sqrt n\}$.
\end{itemize}
Begin with $A_3$,
\begin{eqnarray}
\lefteqn{\int_{A_3}\left | {\int \pi\left(\hat{\theta}_2+\frac{t}{\sqrt n}\right)\exp\big[w(t)\big]}\Pi(dg|\mathbb X_n)-
{e^{\frac{-t^2h_n}{2}}\pi(\hat{\theta}_2)}  \right |dt}\nn\\
&\leq&\int_{A_3}\left|  {\int \pi\left(\Hat{\theta}_2+\frac{t}{\sqrt n}\right)\exp\big[w(t)\big]}\Pi(g|\mathbb X_n) \right|dt+\int_{A_3}\left|  {e^{\frac{-t^2h_n}{2}}\pi(\hat{\theta}_2)}  \right|dt.\nn
\end{eqnarray}
The first integral goes to $0$ by Condition (B1). The second goes to 0 by the tail property of normal distribution.

Because $\hat{\theta}_2\to \theta_0$, by Taylor expansion, for large n,
\begin{eqnarray}
\lefteqn{w(t)}\nn\\
\!&=&\!\!\!\!-2n\int \left [\dot D_H(\hat {\theta}_2,g)\frac{t}{\sqrt n}+\ddot D_H(\hat {\theta}_2,g)\frac{t^2}{2 n}+\dddot D_H( {\theta}',g)\frac{t^3}{6\sqrt n^3}\right ]\Pi(dg|\mathbb X_n)\nn\\
&\!=&\!\!\!\!\frac{t^2}{2}h_n+R_n
\end{eqnarray}
for some  $\theta'\in (\theta_0,\hat{\theta}_2)$. Now consider
\begin{eqnarray}
\lefteqn{\int_{A_1}\left | \int{ \pi\left(\hat{\theta}_2+\frac{t}{\sqrt n}\right)\exp\big[w(t)\big]}\Pi(dg|\mathbb X_n)-
{e^{\frac{-t^2h_n}{2}}\pi(\hat{\theta}_2)}  \right |dt}\nn\\
&\to&
\int_{A_1}\left| { \pi\left(\hat{\theta}_2+\frac{t}{\sqrt n}\right)\exp\left[-\frac{t^2I(\theta_0)}{2}+R_{n0}\right]}-
{e^{\frac{-t^2I(\theta_0)}{2}}\pi(\hat{\theta}_2)}  \right|dt\nn\\
&\leq&  \int_{A_1}\pi(\hat{\theta}_2)\left| e^{\frac{-t^2I(\theta_0)}{2}+R_{n0}} - e^{\frac{-t^2I(\theta_0)}{2}}   \right|dt\nn\\
&&+
\int_{A_1}\left| \pi\left(\hat{\theta}_2+\frac{t}{\sqrt n}\right)- \pi(\hat{\theta}_2) \right|e^{\frac{-t^2I(\theta_0)}{2}}dt \nn
\end{eqnarray}
The second integral goes to $0$ in $P_{f_{\theta_0}}$ probability since $\pi$ is continuous at $\theta_0$. The first integral equals
\begin{eqnarray}\label{eq:aa1}
\int_{A_1}\pi(\hat{\theta}_2)e^{\frac{-t^2I(\theta_0)}{2}}\left| e^{R_{n0}}-1   \right|dt
\leq
\int_{A_1}\pi(\hat{\theta}_2)e^{\frac{-t^2I(\theta_0)}{2}} e^{|R_{n0}|}| R_{n0}|dt
\end{eqnarray}
Now,
$$
\sup_{A_1}R_{n0}=\sup_{A_1}\left(\frac{t}{\sqrt n}\right)^3\dddot D_H(\theta',g_0)\leq c^3\frac{(\log n)^3}{\sqrt n}O_p(1)=o_p(1)
$$
 and hence (\ref{eq:aa1}) is
 $$
 \leq \sup_{A_1} \pi\left(\hat{\theta}_n+\frac{t}{\sqrt n}\right)\int _{A_1}e^{\frac{-t^2I(\theta_0)}{2}} e^{|R_{n0}|}| R_{n0}|dt=o_p(1).
 $$
Next consider
\begin{eqnarray}
\lefteqn{\int_{A_2}\left |\int { \pi\left(\hat{\theta}_2+\frac{t}{\sqrt n}\right)\exp\big[w(t)\big]}\Pi(dg|\mathbb X_n)-
{e^{\frac{-t^2h_n}{2}}\pi(\hat{\theta}_2)}  \right |dt}\nn\\
&\to&
\int_{A_2}\left| { \pi\left(\hat{\theta}_2+\frac{t}{\sqrt n}\right)\exp\left[-\frac{t^2I(\theta_0)}{2}+R_{n0}\right]}-
{e^{\frac{-t^2I(\theta_0)}{2}}\pi(\hat{\theta}_2)}  \right|dt\nn\\
&\leq& \int_{A_2}  \pi\left(\hat{\theta}_2+\frac{t}{\sqrt n}\right)\exp\left[-\frac{t^2I(\theta_0)}{2}+R_{n0}\right]dt+
\int _{A_2}{e^{\frac{-t^2I(\theta_0)}{2}}\pi(\hat{\theta}_2)}  dt. \nn
\end{eqnarray}
The second integral above is
$$
\leq 2\pi(\hat{\theta}_2)e^{\frac{I(\theta_0)c\log \sqrt n}{2}}[\delta\sqrt n-c \log \sqrt n]\lesssim \pi(\hat{\theta}_2)\frac{\sqrt n}{n^{cI(\theta_0)/4}},
$$
so that by choosing $c$ large, the integral goes to $0$ in $P_{f_{\theta_0}}$ probability.

For the first integral, because $t\in A_2$ and $c\log \sqrt n <|t| <\delta \sqrt n$, we have that $|t|/\sqrt n<\delta$. Thus $R_{n0}\leq \frac{\delta t^2}{6n}\dddot D_H(\theta_0,g_0)$.

Since $\sup_{\theta'\in (\theta_0-\delta,\theta_0+\delta)}\frac{\dddot D_H(\theta')}{n}$ is $O_p(1)$, by choosing $\delta$ small we can ensure that for any $\epsilon$ there is $n_0$ depending on $\epsilon$ and $\delta$ such that
$$
P_{f_{\theta_0}}\left\{ |R_{n0}|<\frac{t^2}{4}h_n,\  \forall t\in A_2  \right\}>1-\epsilon \mbox{ for } n>n_0.
$$
Hence, with probability greater than $1-\epsilon$,
$$
\int_{A_2}  \pi(\Hat{\theta}_2+\frac{t}{\sqrt n})e^{-\frac{t^2I(\theta_0)}{2}+R_{n0}}dt\leq \sup_{t\in A_2}\pi(\hat{\theta}_2+t/\sqrt n)\int_{A_2}e^{-t^2I(\theta_0)/4}dt\to 0
$$
as $n\to \infty$.
Combining the three parts together by choosing $\delta$ for $A_2$, and then applying this $\delta$ to $A_1$ and $A_3$ completes the proof.
$\Box$

\end{document}